\documentclass[twocolumn,showpacs,pra]{revtex4}
\usepackage{graphicx}
\usepackage{color}
\usepackage{amsfonts}
\usepackage{amsmath}

\newcommand{\pd}[2]{\frac{\partial #1}{\partial #2}}

\newcommand{\br}[1]{\left(#1\right)}
\newcommand{\fracb}[2]{\br{\frac{#1}{#2}}}
\newcommand{\na}{\varphi_N}
\newcommand{\finv}{f^{-1}}
\newcommand{\im}{\mathrm{Im}}
\newcommand{\re}{\mathrm{Re}}

\newcommand{\nvec}{\mathbf{n}}
\newcommand{\ee}{\mathrm{e}}
\renewcommand{\i}{\mathrm{i}}
\newcommand{\eiph}{\ee^{\i \varphi}}
\newcommand{\aspectsharp}{\varepsilon_s}
\newcommand{\aspectround}{\varepsilon_r}

\begin{document}

\title{Texture and shape of two-dimensional domains of nematic liquid crystal}

\author{R. M. W. van Bijnen$^{1}$, R. H. J. Otten$^{1,2}$, and P. van der Schoot$^{1,3}$}
\address{$^{1}$ Department of Applied Physics, Eindhoven University of Technology, P.O. Box 513, 5600 MB Eindhoven, The Netherlands}
\address{$^{2}$ Dutch Polymer Institute, P.O. Box 902, 5600 AX Eindhoven, The Netherlands}
\address{$^{3}$ Institute for Theoretical Physics, Utrecht University, Leuvenlaan 4, 3584 CE Utrecht, The Netherlands}

\begin{abstract}
We present a generalized approach to compute the shape and internal structure of two-dimensional nematic domains. By using conformal mappings, we are able to compute the director field for a given domain shape that we choose from a rich class, which includes drops with large and small aspect ratios, and sharp domain tips as well as smooth ones. Results are assembled in a phase diagram that for given domain size, surface tension, anchoring strength, and elastic constant shows the transitions from a homogeneous to a bipolar director field, from circular to elongated droplets, and from sharp to smooth domain tips. We find a previously unaccounted regime, where the drop is nearly circular, the director field bipolar and the tip rounded. We also find that bicircular director fields, with foci that lie outside the domain, provide a remarkably accurate description of the optimal director field for a large range of values of the various shape parameters.
\end{abstract}


\maketitle

\begin{section}{Introduction\label{SecIntroduction}}
Nematic liquid crystals are uniaxial fluids that respond elastically to an externally imposed deformation of the director, i.e., the locally preferred orientation of the molecules. The interfacial properties of nematic liquid crystals are also anisotropic. This means that the surface free energy depends on the attack angle of the director relative to the interface \cite{degennes93,prinsen03,prinsen04,fraden85,drzaic95,tang95,riviere95,rapini69}. Under conditions of spatial confinement, i.e., in a cavity, surface and bulk elastic properties compete so as to minimize the overall free energy \cite{burylov97,heras09,ondris93}.

For instance, homeotropic anchoring of the liquid crystal to the interface in a spherical drop or cavity may lead to a radial director field, whereas planar anchoring produces a bipolar field \cite{williams86}. The former is characterized by a hedgehog point defect in the center \cite{verhoeff09,verhoeff11lm,verhoeff11jcp}, and the latter by two surface point defects called boojums \cite{rudnick95,rudnick99}. If the confinement is soft, which happens to be the case if the nematic droplet is suspended in a fluid \cite{zocher29}, not only the director field adjusts to the presence of the interface, but the interface itself adjusts to the director field as well \cite{bernal41}.

Spindle-shaped nematic droplets, or tactoids, have indeed been observed in suspensions of elongated colloidal particles, under conditions where isotropic and nematic phases coexist \cite{kaznacheev02}. Their shape and internal structure has been successfully described in terms of the bulk elastic and surface properties by applying variational theories \cite{kaznacheev02, prinsen03, prinsen04}. These make use of plausible families of droplet shape and director-field structure, and have been applied to extract information on anchoring strengths and elastic constants from polarization microscopic images of tactoids \cite{prinsen03, prinsen04, kaznacheev02,verhoeff09,verhoeff11lm,verhoeff11jcp}.

Interestingly, values of the ratio of the anchoring strength and the surface tension obtained by curve fitting to experiments were found to be very much larger than expected from theoretical prediction \cite{prinsen03, prinsen04, puech10}. It is not clear if this discrepancy is caused by the variational nature of the theories, warranting a more precise theoretical investigation of the coupling between director field and droplet shape. In this paper we do just that, by invoking Frank elasticity theory and a Rapini-Papoular-type description of the interfacial properties of the nematic \cite{frank58, rapini69}. We restrict ourselves to the two-dimensional case, of which physical realizations may be found in certain types of Langmuir monolayer \cite{rudnick95,fang97,rudnick99,loh98,fischer94,pettey99,riviere95,schwartz94}.

For this kind of two-dimensional system, and in the so-called equal constant approximation, we develop a fast, and easily implementable numerical method for finding the optimal director field for \textit{fixed} arbitrary shapes and arbitrary interfacial energies. Our method takes the form of a simple iterative scheme and employs conformal mappings and fast fourier transforms. Hence, the method is useful for studies of nematics in confined geometries \cite{drzaic95,Burylov97, davidson11}. However, we are also able to apply our method to study properties of two-dimensional droplet domains that are soft and can optimize their shape. Our method provides an alternative to numerically more involved finite element studies \cite{loh98, loh00}.

As opposed to earlier quasi-analytical studies \cite{prinsen03, prinsen04, kaznacheev02, kaznacheev03}, which need to variationally optimize both the director fields and domain shapes within certain classes, in our case there remains only one parameter to be optimized, being the shape of the domain. We pick a rich class of physically plausible spindle shapes, which consist of a domain bounded by circle sections, and extend it to include smooth, rounded tips. Within this class of shape, we compute a state diagram mapping out the director field and the domain shape, such as the tip being round or sharp, as a function of the various material constants and domain size. In addition, we show that for the spindle shapes with sharp tips, the equilibrium director fields are accurately described by fields constructed from circle sections.

In section \ref{SecFreeEnergy}, we first describe the free energy of our model and discuss the relevant dimensionless groups describing the physics of the problem. Section \ref{SecFixedDomain} discusses the equations that determine the optimal directorfield for fixed domains, and our light-weight numerical method for finding these optimal fields. In section \ref{SecSpindle}, we put forward a parametrization of a class of domain shapes that interpolate smoothly between sharp and rounded spindle domains. In section \ref{SecFixSpindle} we present optimal director fields calculated for the mentioned class of droplet shapes. Finally, in section \ref{SecOptimalShape}, we variationally determine the optimal shape within our class of domain, and present a state diagram of optimal droplet shapes and director fields as a function of the relevant material parameters. We end the paper with a summary of our findings and a discussion of possible future extensions of our work in section \ref{SecConclusions}.
\end{section} % Introduction

\begin{section}{Free energy \label{SecFreeEnergy}}
In order to describe the domain shapes and director fields, we use a macroscopic theory based on a competition between the elastic and surface properties of nematic droplets that float freely in an isotropic phase. We model this by applying the Frank elasticity theory and a suitable anchoring surface energy that couples the director field to the interface, and restrict ourselves to two spatial dimensions. 

The director field ${\bf n}$ indicates the average orientation of the particles in the nematic phase and we write it as
\begin{equation}\label{OrderParameter}
{\bf n} = (\cos \Theta, \sin \Theta),
\end{equation}
where $\Theta=\Theta(x,y)$ is the angle that a particle at position $(x,y)$ makes with the horizontal axis, which we choose as the axis of mirror symmetry. The free energy $F=F_{e}+F_{s}$ of a nematic drop of given surface area $S$ consists of two contributions: an elastic contribution $F_e$ and a surface (boundary) contribution $F_s$:

\textit{i)} The Oseen-Frank elastic energy $F_e$ is associated with the distortion of the  director field. In our two-dimensional model the deformation energy density comprises only a splay and a bend contribution \cite{frank58},
\begin{equation}\label{ElasticEnergy}
F_e = \frac{1}{2}\int_\Omega \left( K_1(\nabla\cdot{\bf n})^2 + K_3(\nabla \times {\bf n})^2\right) dA,
\end{equation}
where $K_1$ and $K_3$ are the splay and bend elastic constants, respectively, and  where the integration is taken over the entire surface of the domain $\Omega$. We adopt the equal-constant approximation, where the splay and bend elastic constants are presumed to be equal, the sum of the elastic deformation energy densities can be written as $K|\nabla \Theta|^2/2$, with $K$ the average elastic constant. 

\textit{ii)} The interfacial energy $F_s$ is associated with the anchoring of the director field to the interface (the phase boundary). It depends on the angle $\na - \Theta$ between the director field $\Theta$, and the angle $\varphi_N$ of the surface normal with the horizontal axis, and in the most general form it is written as
\begin{equation}\label{BoundaryEnergy}
F_s = \int_{\partial\Omega}\sigma(\na - \Theta) dS,
\end{equation}
where $\partial \Omega$ is the boundary of the domain, and the energy density $\sigma$ is given by
\begin{equation}\label{BoundaryEnergyGeneral}
\sigma(\na - \Theta) = a_0 + \sum_{n=1}^\infty a_n \cos n(\na - \Theta).
\end{equation}

Here, $a_0$ is the bare surface (line) tension that specifies the energy per unit length of the boundary, irrespective of the director field $\Theta$. The higher order $a_n$ specify how the boundary energy depends on the director field orientation. Rudnick and co-workers studied the case of only a single $a_n$ unequal to zero \cite{rudnick95,rudnick99} and nonzero values of both $a_1$ and $a_2$ \cite{loh98, loh00}. If we retain only $n=2$ in the sum in Eq. (\ref{BoundaryEnergyGeneral}), corresponding to the interfacial energy of a two-dimensional nematic \cite{rudnick95}.  In that case, the interfacial energy density becomes of the Rapini-Papoular type \cite{rapini69}
\begin{equation}\label{BoundaryEnergya0a2}
\sigma(\na - \Theta) = a_0 + a_2 \cos 2(\na - \Theta),
\end{equation}
where $a_2$ is the anchoring strength of the director field. Note that the $a_2$-term can be both positive and negative, so in order to prevent unphysical negative total surface energy densities, we must have the restriction that $|a_2| \le a_0$. 

From now on, we restrict ourselves to the case of planar alignment, implying that the director field favours a tangential orientation to the boundary. However, we shall point out that fields with a preference for homeotropic anchoring can be easily obtained with our methodology as well.

Combining the above two energy contributions, we then have for the total free energy
\begin{equation}\label{FreeEnergy}
F[\Theta] = \frac{K}{2}\int_{\Omega} |\nabla \Theta|^2 dA + \int_{\partial\Omega}\sigma(\na - \Theta)dS.
\end{equation}
We will work with a fixed domain area $A_0$ and all lengths are scaled in units of
\begin{equation}\label{EqTypicalLength}
\rho = \sqrt{A_0/\pi}.
\end{equation}
We make the bare surface (line) tension $a_0$ and anchoring strength $a_2$ dimensionless by introducing
\begin{equation}\label{EqOmega}
\omega \equiv \frac{2a_0 \rho}{K},
\end{equation}
and
\begin{equation}\label{EqGamma}
\gamma \equiv \frac{2a_2 \rho}{K}.
\end{equation}
Note that in view of the restriction $0 < a_2 \leq a_0$ we always have that $\gamma \leq \omega$. We can write $\gamma= 2\rho/\xi$ with $\xi \equiv K/a_2$ the extrapolation length \cite{prinsen03}, so small values of $\gamma$ should give rise to a homogeneous director field, whereas a curved, bipolar field is to be expected for large values of $\gamma$.

In order to obtain both the equilibrium shape \emph{and} director field of a domain, one should simultaneously minimize the total free energy Eq. (\ref{FreeEnergy}) with respect to $\Theta$ and $\Omega$. However, this is a formidable mathematical problem that is not easily solved analytically  \cite{rudnick95, kalugin98}, and still difficult numerically \cite{loh98, loh00}. Therefore, we first consider the problem of a fixed shape $\Omega$ in the next section, and develop an easy-to-implement, numerical method for finding the optimal director field that minimizes the free energy (\ref{FreeEnergy}).  

\end{section}

\begin{section}{Calculating the optimal texture of a fixed domain\label{SecFixedDomain}}
If the shape of the domain $\Omega$ is kept fixed, variational minimization of the free energy (\ref{FreeEnergy}) with respect to the director field $\Theta$ leads to the following boundary value problem for $\Theta$ \cite{rudnick95}:
\begin{eqnarray}
\nabla^2 \Theta(x,y) &=& 0, \hspace{1cm} (x,y) \in \Omega \label{Laplace}\\
& &  \nonumber \\
\kappa \pd{\Theta(x,y)}{n} &=& g[\Theta](x,y), \hspace{1cm} (x,y)\in \partial\Omega,\label{dTdn}
\end{eqnarray}
with
\begin{equation}\label{gTheta}
g[\Theta](x,y) = \sigma'(\na(x,y) - \Theta(x,y)).
\end{equation}
Here, $\na$ is the angle that the normal to the boundary makes with the reference axis, and $\sigma'(u) = d\sigma/du$ with $\sigma$ the interfacial free energy density (a free energy per unit length), Eq. (\ref{BoundaryEnergyGeneral}).
For arbitrary domains $\Omega$, the nonlinear boundary condition makes the boundary value problem posed by Eqs. (\ref{Laplace}) and (\ref{dTdn}) highly intractable for solution by analytical means. We will therefore resort to numerics: using conformal mappings we reduce the problem to a single integral equation for the director field $\Theta$ on the boundary of the domain, which is solved by an iterative procedure employing Fourier transforms. This method is fast, easy to implement, and suited for arbitrary domain shapes. In the remainder of this section we will outline our solution procedure.

Inspired by the work of Rudnick and collaborators \cite{rudnick95}, we start by using the complex variable $z = x + \mathrm{i} y$ to represent the two dimensional coordinate $(x,y) \in \Omega$, thus gaining access to the powerful methods of the field of complex analysis. In particular, we use conformal mappings to map the domain $\Omega$ onto the unit disk $D$. A conformal map is an analytic function that maps one domain onto another, while preserving harmonicity of functions defined on either domain \cite{Nehari}.

Let
\begin{equation}
w = f(z)
\end{equation}
be an analytic function that maps the domain $\Omega$ conformally onto the unit disk $D: |w| < 1$ and $\partial \Omega$ onto $\partial D$. The process of mapping coordinates $z$ of the domain $\Omega$ conformally onto coordinates $w$ of the unit disk $D$ is depicted schematically in Fig. \ref{FigConfmap}.
\begin{figure}[h]
\includegraphics[width=0.85\columnwidth]{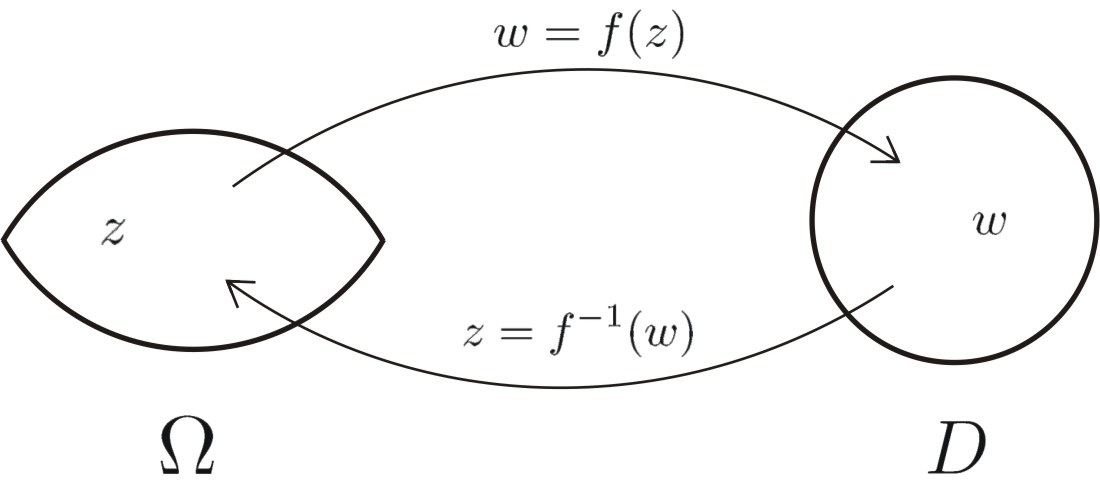}
\caption{The conformal map $f: \mathbb{C}\to \mathbb{C}$ maps a point $z \in \Omega$ bijectively to a point $w \in D$, the unit disk. The inverse $f^{-1}$ performs the opposite mapping.}  \label{FigConfmap}
\end{figure}
If, and only if, $\Theta$ is harmonic on $\Omega$, then the function
\begin{equation}
\Xi(w) = \Theta(\finv(w)), \hspace{1cm} w\in D
\end{equation}
is harmonic on $D$, where $f^{-1}$ denotes the inverse operation of $f$. Moreover, if $\Theta$ satisfies the boundary condition (\ref{dTdn}), then $\Xi$ satisfies the modified boundary condition on $\partial D$ \cite{Henrici, Churchill, Dettman}:
\begin{equation}\label{hXi}
\pd{\Xi}{n_D} = \frac{g[\Xi \circ f](f^{-1}(w))}{|f'(f^{-1}(w))|} \equiv h[\Xi](w),
\end{equation}
where $\pd{}{n_D}$ denotes the normal derivative on the unit disk $D$, and the functional $g[\cdot](\cdot)$ is defined in (\ref{gTheta}).
It follows that the problem (\ref{Laplace}, \ref{dTdn}) on $\Omega$ can be mapped onto a Neumann problem on the unit disk $D$:
\begin{equation}\label{NMXi}
\left \{
\begin{array}{rclr}
				\nabla^2 \Xi &=& 0 & \ \ \ \ \ (w \in D) \\
				\ & \ & \ & \ \\
				\dfrac{\partial \Xi}{\partial n_D} &=& h[\Xi](w) & \ \ \ \ \
(w \in \partial D) \\
\end{array} \right.
\end{equation}
with $h$ defined as in Eq. (\ref{hXi}).
Once the solution $\Xi$ to this problem is found, the solution $\Theta$ to the original problem (\ref{Laplace}), (\ref{dTdn}) can easily be reconstructed as $\Theta = \Xi \circ f$.

We solve the above problem by iteration. Starting with an initial guess $\Xi_0$, successive approximations $\Xi_{n+1}$ ($n = 0, 1, \ldots$) to the solution of (\ref{NMXi}) are obtained by solving the intermediate Neumann problem
\begin{equation}\label{NMXin}
\left \{
\begin{array}{rclr}
				\nabla^2 \Xi_{n+1} &=& 0 & \ \ \ \ \ (w \in D) \\
				\ & \ & \ & \ \\
				\dfrac{\partial \Xi_{n+1}}{\partial n_D} &=& h[\Xi_n](w) & \ \ \ \ \
(w \in \partial D), \\
\end{array} \right.
\end{equation}
where it should be noted that the boundary condition is now \textit{linear}. This problem is an ordinary Neumann problem, which is solved by \cite{Henrici}
\begin{equation}\label{intdisk}
\Xi_{n+1}(w) = -\frac{1}{2 \pi}\int_{\partial D} \mathcal{N}(w, w') h[\Xi_n](w') \mathrm{d}w',
\end{equation}
where the \textit{Neumann function} $\mathcal{N}(w, w')$ is specified by
\begin{equation}\label{NeumannF}
\mathcal{N}(w, w') = -\log \br{|w - w'||1 - \overline{w}w'|},
\end{equation}
and where $\overline{w}$ denotes the complex conjugate of $w$.

The nonlinear integral Eq. (\ref{intdisk}) provides an iteration scheme for constructing successive approximations $\Xi_{n+1}$ of the director field, based only on its values on the boundary of the unit disk. It suffices therefore to only calculate the director field values on the boundary throughout the iteration process. Should iteration of the integral Eq. (\ref{intdisk}) yield a converging sequence $\Xi_{n}$ for $n\rightarrow\infty$, then we have in principle solved problem (\ref{NMXi}).

However, evaluating the integral numerically at each iteration step for a discrete set of boundary points turns out to be a cumbersome and slow process. Instead, the integration step is circumvented by considering a Fourier series expansion of $h[\Xi_n](w)$, with $w = \exp(\mathrm{i} \varphi) \in \partial D$:
\begin{equation}\label{FourierExpansion}
h[\Xi_n](\eiph) = \sum_{k = 1}^\infty b_k^{(n)} \sin k \varphi.
\end{equation}
Because of the nematic symmetry of the director field and domain $\Omega$, which is respected by the conformal map $f$, we only need to consider sinusoidal terms. The Fourier coefficients $b_k$, of a function $h(\varphi)$, can be obtained from the forward Fourier sine transform $\mathcal{F}_k$, which we define as
\begin{equation}
b_k = \mathcal{F}_k[h(\varphi)] \equiv \frac{1}{\pi}\int_0^{2\pi} h(\varphi) \sin k \varphi \mathrm{d} \varphi,
\end{equation}
and similarly we define the inverse Fourier sine transform $\mathcal{F}^{-1}_\varphi$, that reconstructs the function $h(\varphi)$ from its Fourier coefficients $b_k$ again:
\begin{equation}
h(\varphi) = \mathcal{F}^{-1}_\varphi[b_k] = \sum_{k = 1}^{\infty} b_k \sin k \varphi.
\end{equation}
Inserting the expansion (\ref{FourierExpansion}) in the integral Eq. (\ref{intdisk}), and noting that when $w, w'\in \partial D$ the Neumann function (\ref{NeumannF}) assumes the particularly simple form
\begin{equation}
\mathcal{N}(\eiph, \ee^{\i \varphi'}) = -\log \br{2 - 2 \cos (\varphi - \varphi')},
\end{equation}
then the integral in Eq. (\ref{intdisk}) can be evaluated analytically \cite{Gradshteyn} to yield
\begin{equation}\label{Xinew}
\Xi_{n+1}(\eiph) = \sum_{k = 1}^\infty \frac{b_k^{(n)}}{k} \sin k \varphi = \mathcal{F}^{-1}_\varphi[b_k^{(n)} / k].
\end{equation}

Using the above calculations and definitions, the successive approximation iteration scheme for the integral Eq. (\ref{intdisk}) turns into an iteration scheme for the expansion coefficients $b_k$:
\begin{equation}\label{itscheme}
b^{(n+1)}_{k} = \mathcal{F}_k\left[\frac{\gamma \sin\br{2 \theta - 2\mathcal{F}_\varphi^{-1}\left[b^{(n)}_{{k}'}/ k'\right]}}{|f'(f^{-1}(\eiph))|^{-1} } \right],
\end{equation}
where $\theta$ is the angle that the normal to $\partial \Omega$ makes with respect to the $x$-axis, evaluated at the point $f^{-1}(\mathrm{e}^{\mathrm{i}\varphi})$. Finding the optimal director field has now been reduced to an iteration of the above equation, and when the sequence converges the director field can simply be found by first calculating $\Xi$ through Eq. (\ref{Xinew}), and subsequently evaluating $\Theta = \Xi \circ f$.
Each iteration step in the above scheme requires only two Fourier transforms, which can be done both accurately and fast using Fast Fourier Transforms (FFTs) \cite{Brigham}. Moreover, the above iteration scheme is easy to implement, making the determination of the optimal director field for a fixed shape with a known conformal map a simple computational task.

We conclude this section with some notes on practical issues. Firstly, the iteration scheme (\ref{itscheme}) is adjusted slightly in order to aid convergence. Rather than replacing an approximation $b_k^{(n)}$ in its entirety by the right-hand side of Eq. (\ref{itscheme}), the new approximation $b_k^{(n + 1)}$ is then composed as a combination of $\mu$ times the right hand side of Eq. (\ref{itscheme}) and $(\mu - 1)$ times the previous approximation $b_k^{(n)}$, with some parameter $0 < \mu < 1$. The convergence of this sequence is then accelerated using a Shanks transformation \cite{BenderOrszag}, which roughly doubles the speed of convergence.

Secondly, when we obtained a solution $\Xi_{n+1}$ to the intermediate Neumann problem (\ref{NMXin}) at iteration step $n$, we have in fact obtained a whole family of solutions. Additional solutions can be found by simply adding a constant $c \in [0, 2 \pi]$, for $\Xi_{n+1} + c$ still solves problem (\ref{NMXin}). The constant $c$ determines the average value of the director field taken over the entire domain. However, when the sequence $\Xi_n$ has converged to some final value $\Xi_\infty$, the same does not hold true for the original problem (\ref{NMXi}).
For if $\Xi_\infty$ solves (\ref{NMXi}), $\Xi_\infty + c$ does in general \textit{not} solve that same problem due to the nonlinearity of the boundary condition.  But problem (\ref{NMXi}) in fact does allow for multiple solutions. Solution (\ref{intdisk}) represents a director field which has an average orientation of $\langle\Xi\rangle=0$, so, in other words, we have tacitly picked the $c=0$ solutions.

To find director fields $\Xi$ with a different average orientation $\langle\Xi\rangle = c$, we need to select the corresponding solution with offset $c$ at each and every step of the iteration.
For domains $\Omega$ with symmetry along the $x$ and $y$ axis, the only two possible values of $c$ are $c = 0$ and $c = \pi/2$ (modulo $\pi/2$). This can be seen from the Neumann condition $\int_{\partial D}h[\Xi](w)\mathrm{d}w = 0$ that has to be satisfied for problem (\ref{NMXi}) to have a solution \cite{Roach}. For elongated domains along the $x$-axis it turns out that  with $a_2>0$ the $c=0$ solution corresponds to an energy minimum and $c = \pi/2$ to a maximum. When $a_2 < 0$ the opposite is true and we should select the $c = \pi/2$ solution at each iteration step.

This concludes the description of our methodology. Note that this method is in principle applicable to any domain $\Omega$, and any arbitrary  boundary energy functional $g[\Theta](z)$. However, in this paper we shall restrict ourselves to nematic phases, with boundary energies of the form of Eq. (\ref{BoundaryEnergyGeneral}). To proceed further, we now need to specify the exact shape of the domain $\Omega$.
\end{section}

\begin{section}{Domain shapes\label{SecSpindle}}
We introduce two classes of domain, as well as their parametrization and conformal mapping, for which we will calculate the optimal director field as outlined in Sec. \ref{SecFixedDomain}. In the next sections, we use these classes of domain to approximate the optimal shape.
The first class of domain is that of spindle-shaped domains with sharp tips, whereas the second class contains domains with rounded tips, and these are constructed from the first class, to be discussed below.

The choice for the spindle shapes is inspired by experimentally observed domain shapes in two \cite{riviere95,schwartz94,dolganov07} and three \cite{puech10,zhang06} dimensions, and the optimal shapes found in numerical studies \cite{loh98, loh00}. Sharp-tipped spindle shapes are domains bounded by two identical circular arcs, as depicted in Fig. \ref{FigSpindleShape}. The circular arcs making up the spindle shape have a radius $R_0$, and intersect each other at the point $z = \pm z_1$ (recalling that $z = x + \mathrm{i} y$), located on the $x$-axis. The $y$-axis is intersected at point $\pm z_2$, and provided that $|z_2| \leq |z_1|$, we can define the aspect ratio of the spindle domain as $\aspectsharp = |z_1| / |z_2|$. The spindle shape in Fig. \ref{FigSpindleShape} has an aspect ratio of $\aspectsharp = 2$.
\begin{figure}[t]
\includegraphics[width=0.9\columnwidth]{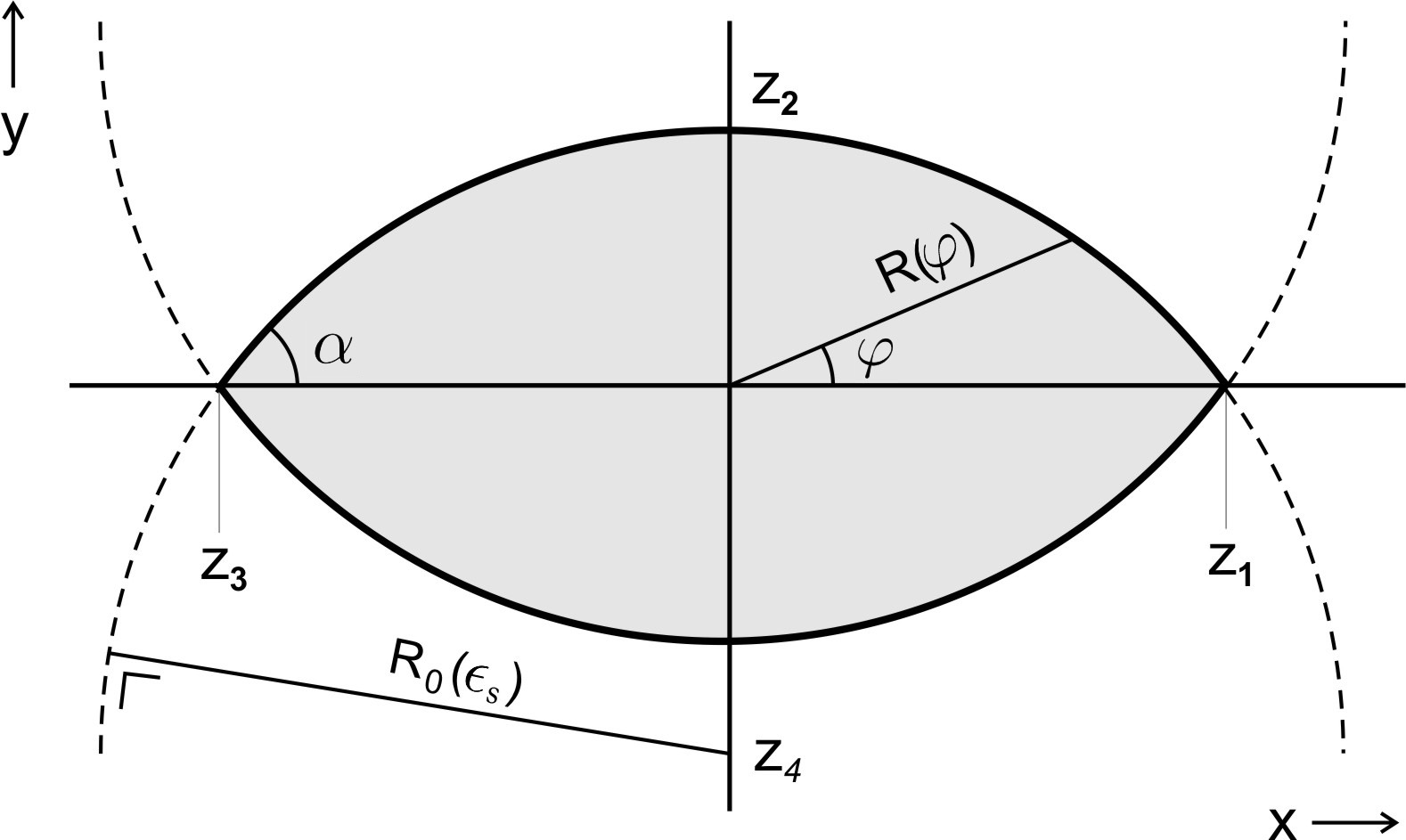}
\caption{Spindle shaped domain with with aspect ratio $\aspectsharp = 2$. Relevant variables in the parametrisation and conformal map are indicated. The symbols $z_1, .., z_4$ that are drawn at a point $(x,y)$, correspond to complex numbers $z = x + \mathrm{i} y$.}  \label{FigSpindleShape}
\end{figure}
Using the area $A_0$ of the domain and aspect ratio $\aspectsharp$ as input parameters, the spindle boundary is parameterized in polar coordinates, in the first quadrant, by
\begin{equation}\label{Rphi}
R(\varphi) = R_0(\aspectsharp) \br{\sqrt{1-\beta^2 \cos^2 \varphi} - \beta \sin \varphi}
\end{equation}
where $\varphi$ is the polar angle, and
\begin{equation}\label{EqBeta}
\beta = \frac{\aspectsharp^2 - 1}{\aspectsharp^2 + 1}.
\end{equation}
The circular arc radius is given by
\begin{equation}
R_0(\aspectsharp) = \sqrt{\frac{A_0}{2\br{\arccos \beta - \beta \sqrt{1-\beta^2}}}}.
\end{equation}
The normal angle to the surface at a point $(R(\varphi), \varphi)$ is equal to
\begin{equation}
\na = \varphi - \arctan \br{\frac{1}{R(\varphi)}\pd{R}{\varphi}}.
\end{equation}
The above parametrizations are restricted to the first quadrant $0 \leq \varphi \leq \pi /2$, but other quadrants can be found by simply mirroring the first quadrant.

The conformal mapping of the spindle shaped domain can be constructed from two consecutive mappings. First, the spindle domain $\Omega$ is mapped to the upper half plane $u \in C, \im(u) \geq 0$ by the conformal map \cite{Kober}
\begin{equation}
u(z) = \br{-\i \frac{(z-z_1)(z_4 - z_3)}{(z-z_3)(z_1 - z_3)}}^{\pi/2\alpha}\hspace{1cm} z \in \Omega,
\end{equation}
with the complex numbers $z_1 = R(0)$, $z_2 = \i R(\pi/2)$, $z_3 = -z_1$ (see Fig. \ref{FigSpindleShape}), and the locus of the upper circle section $z_4 = \i R(\pi/2) - \i R_0(\aspectsharp)$. The angle $2\alpha$ is that of the sharp tip of the spindle, as indicated in Fig. \ref{FigSpindleShape}, and is calculated as
\begin{equation}\label{EqCuspAngle}
\alpha = \arcsin\fracb{z_1}{R_0(\aspectsharp)} = \arccos \beta,
\end{equation}
with $\beta$ as defined in Eq. (\ref{EqBeta}). Next, defining
\begin{equation}
v(u) = \frac{u-u_1}{u-u_2}\frac{w_3 - w_1}{w_3 - w_2},
\end{equation}
and
\begin{equation}
g(u) = \frac{w_1 - v(u)w_2}{1-v(u)},
\end{equation}
then $w = g(u)$ is the conformal map that maps the upper half plane to the unit disk in the $w$-plane, where the points $u_1$, $u_2$ and $\infty$ of a line in the upper half plane are mapped onto the points $w_1 = 1$, $w_2 = \i$ and $w_3 = -1$ on the unit disk. When we pick $u_1 = u(z_1)$ and $u_2 = u(z_2)$, then the composite map (see also Fig. \ref{FigConfmap})
\begin{equation}\label{confmapf}
f(z) = g(u(z))
\end{equation}
maps the spindle domain onto the unit disk, where the $i$-th quadrant in the $z$-plane is mapped to the $i$-th quadrant in the $w$-plane ($i = 1, 2, 3, 4$).

The inverse map, mapping the unit disk conformally to a spindle domain, can be computed by inverting Eq. (\ref{confmapf}), resulting in
\begin{equation}\label{confmapfinv}
\finv(w) = \frac{z_1 + z_2+ \br{-\i\frac{w-1}{w+1}}^{2\alpha/\pi}(z_2 - z_1)}{1+\br{-\i\frac{w-1}{w+1}}^{2\alpha / \pi}\br{1-\frac{z_2}{z_1}} + \frac{z_2}{z_1}}.
\end{equation}

Interestingly, through the introduction of an additional parameter $\lambda$, with $0 < \lambda \leq 1$,  a whole new family of domain shapes $\Omega_\lambda$ can be generated with the conformal mapping
\begin{equation}\label{ConfSmooth}
z_\lambda = A_\lambda \finv(\lambda w),\hspace{1cm} w \in D
\end{equation}
which maps only a part of $D$ onto the $z$-domain, creating a more rounded spindle shape without a sharp tip. The new smoothed spindle shapes are generally smaller than the original spindle shape, and a constant $A_\lambda$ is added for area conservation. This class of shapes transforms fluently from the sharp-tipped spindle domain ($\lambda = 1$) to the unit disk ($\lambda \rightarrow 0$), and they bear strong resemblance to the fully numerically calculated optimal results \cite{loh98, loh00}. The family of smooth spindle shapes is shown in Figure \ref{FigSmoothSpindle}, where we have temporarily set $A_\lambda = 1$ for clarity.
\begin{figure}[t]
\includegraphics[width=0.9\columnwidth]{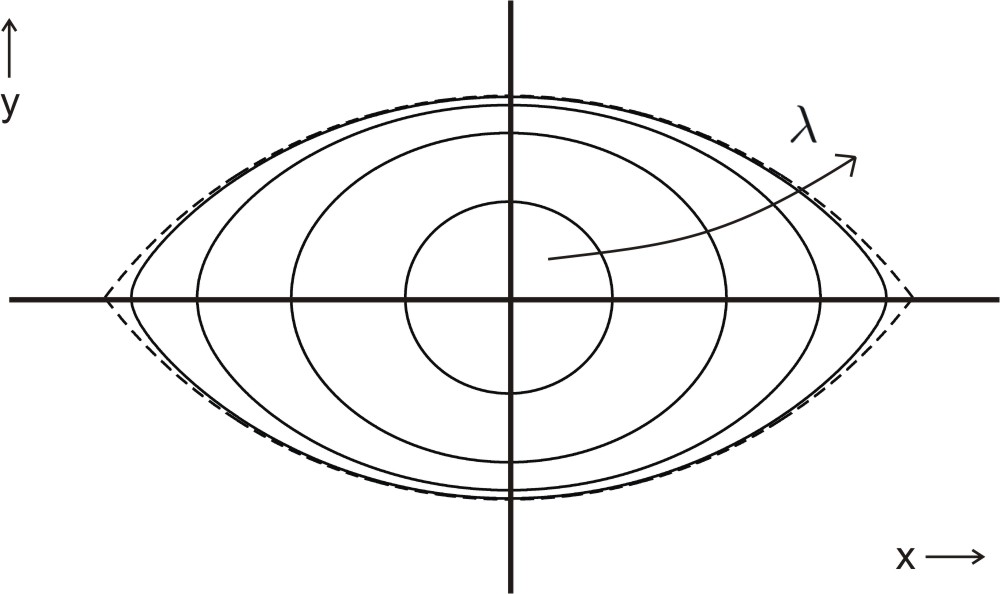}
\caption{The family of smooth spindle shapes generated by the conformal map (\ref{ConfSmooth}), for various values of $\lambda$ between $0$ and $1$ with outward lying shapes for increasing $\lambda$ (indicated by arrow). The outermost shape drawn with a dashed line, is the original, sharp-tipped spindle shape with aspect ratio $\aspectsharp = 2$ and $\lambda = 1$.}  \label{FigSmoothSpindle}
\end{figure}

The proper value of $A_\lambda$ can be calculated by noting that the droplet area $A_0$ can be calculated using the divergence theorem:
\begin{equation}
2 A_0 = \int_{\Omega_\lambda} \nabla \cdot (x,y) dA = \int_{\partial \Omega_\lambda} (x,y)\cdot \nvec dS = A_{\lambda}^2 I_{\lambda},\label{Alambda0}
\end{equation}
with
\begin{equation}
I_\lambda = \int_{\partial D}\left(\re\ \psi(\theta) \im\ \pd{\psi}{\theta} - \im\ \psi(\theta) \re\ \pd{\psi}{\theta}\right) d\theta,
\end{equation}
and where we have written $\psi(\theta) \equiv f^{-1}(\lambda \exp(\i \theta))$. In order to conserve the area of the droplet $A_0$ when varying $\lambda$, we thus must have that
\begin{equation}\label{Alambda}
A_\lambda = \sqrt{\frac{2A_0}{I_\lambda}}.
\end{equation}

The parameters $\aspectsharp$ and $\lambda$ provide not much insight into the physical shape of the domain. Moreover, the droplet shape is very (in)sensitive to small changes in $\lambda$ for $\lambda$ close to $1 (0)$. Therefore, we introduce two new and more convenient descriptors, being the aspect ratio of the round-tipped droplet
\begin{equation}\label{epseff}
\aspectround= \aspectround(\aspectsharp, \lambda) = \frac{f^{-1}(\lambda)}{|f^{-1}(\mathrm{i} \lambda \pi / 2)|},
\end{equation}
and the radius of curvature $R_c$ at the tip of the domain,
\begin{equation}\label{Rc}
R_c(\aspectsharp, \lambda) = \frac{8K A_\lambda \lambda \aspectround \alpha}{\pi (1 + K)^2(1+\lambda)^2+4\alpha \lambda(K^2 - 1)},
\end{equation}
with $2\alpha$ the tip angle as in Eq. (\ref{EqCuspAngle}), and where we have defined
\begin{equation}
K = \fracb{\aspectsharp - \mathrm{i}}{\aspectsharp + \mathrm{i}}^2\fracb{\lambda - 1}{\lambda + 1}^{2\alpha/\pi}.
\end{equation}
The radius of curvature as defined above is defined in units of the waist size $|f^{-1}(\mathrm{i} \lambda \pi / 2)|$ of the droplet, such that it is always a number between $0$ and $1$. Also, in the case of $\lambda = 1$, we have that $\aspectround = \aspectsharp$, and in the remainder of this paper we shall use $\aspectround$ to indicate the aspect ratio of both round-tipped and sharp-tipped shapes.

Having introduced the class of shapes of interest, it becomes important to note an implementation issue for the solution method of Sec. \ref{SecFixedDomain} for elongated domains $\Omega$ with sharp tips, such as spindle shapes with high aspect ratios. Conformal mappings between an elongated domain and the unit disk suffer from a phenomenon called \textit{crowding} \cite{Wegmann, DeLillo}. The concept of crowding can be understood as follows. Consider a set of uniformly distributed points $z \in \Omega$. When $\Omega$ is highly elongated, the density of images $w = f(z)$ of these points will become highly non-uniform in the unit disk. Conversely, a uniformly spaced grid on the unit disk gets mapped to a highly non-uniform spaced grid in the domain $\Omega$. The number of gridpoints on the unit disk required to maintain a certain density of images on $\Omega$ scales exponentially with the aspect ratio of the domain \cite{Wegmann}. 

Additional crowding originates from highly distorted boundary curves $\partial \Omega$ \cite{Zemach}, as occurs for instance in the case of a sharp tip in the domain. For the spindle shaped domains discussed in the next section, our solution method, which requires equidistant grid points on $\partial D$ on behalf of the FFT, needs in the order of up to $10^6$ gridpoints to correctly sample the director field in the tip of the domain when the aspect ratio of the droplet domain is valued around $3$. Due to the efficiency of the FFTs such large grids do not pose a problem in terms of computation time, but the exponential scaling of the number of required gridpoints does prevent us from accurately investigating higher aspect ratios than 3.

\end{section}

\begin{section}{Optimal director field for fixed spindle shapes}\label{SecFixSpindle}

In this section we employ our solution procedure of Sec. \ref{SecFixedDomain}, to calculate the optimal director field for a fixed shape from the spindle classes introduced in the previous section. Before turning to the general case, it is interesting to consider the simplest spindle shape: a circular domain. For this domain the conformal mapping is simply the identity operation. Despite the complicated nonlinear boundary condition in Eq. (\ref{gTheta}), Rudnick and Bruinsma \cite{rudnick95} found an exact analytical solution on the unit disk $D$, provided $a_0$ and only one other $a_k$ in the expansion (\ref{BoundaryEnergyGeneral}) is nonzero. Their solution is of the form
\begin{equation}
\Theta(x, y) = \frac{1}{\i}(q(x+\i y) - q(x - \i y)),
\end{equation}
with
\begin{equation}
q(z) = \frac{1}{k}\log\br{1 - \frac{z^k}{L^k}},
\end{equation}
and the length $L$ given by
\begin{equation}\label{Ldisk}
L^2 = \frac{1 + \sqrt{1+\gamma^2}}{\gamma}.
\end{equation}

In the case of a nematic liquid crystalline phase, for which $k=2$, it can be shown  that the above director field is equal to a bicircular field that has field lines consisting of circular sections intersecting the $x$-axis at points $\pm L$, as illustrated in Fig. \ref{FigBicircularField}. These virtual defects lie outside the domain $\Omega$, and are commonly referred to as \textit{boojums} in the limiting case for which they reside exactly on the boundary of the domain. 

The representation of such a bicircular field in polar coordinates $(r, \varphi)$ is given by
\begin{equation}
\Theta(r, \varphi) = \arctan \frac{-2r^2 \cos\varphi \sin \varphi}{L^2 + r^2\br{\sin^2\varphi - \cos^2 \varphi}}.
\end{equation}
As we shall see, this field is a very good approximation to the solution of the problem (\ref{Laplace}), (\ref{dTdn}) if the domain $\Omega$ is spindle shaped, to which we shall now turn our attention.
\begin{figure}[h]
\includegraphics[width=0.9\columnwidth]{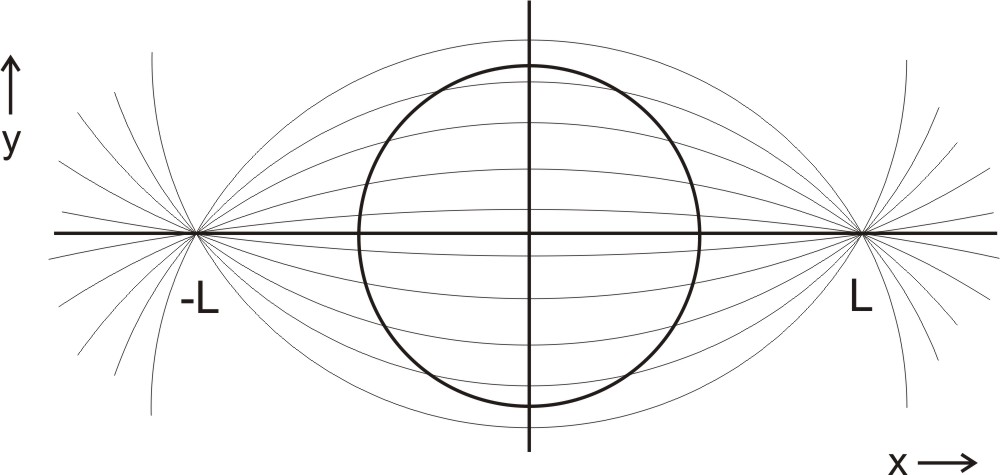}
\caption{The field lines of a bicircular field are circle sections, intersecting the $x$-axis at points $\pm L$. Bicircular fields correspond to the lowest energy configuration of the director field for circular droplet domains.}  \label{FigBicircularField}
\end{figure}

Starting with spindle shapes with sharp tips, a typical example solution of the problem (\ref{Laplace}), (\ref{dTdn}), is shown in Fig. \ref{FigFieldLines}(a) by means of a plot of director field lines. Here, the anisotropic line tension (anchoring strength) is set equal to $\gamma = 2$ [see Eq. (\ref{EqGamma})], and the domain aspect ratio fixed at $\aspectsharp = 2$. Figure \ref{FigFieldLines}(b) shows the optimal director field for the same parameters, but with a round-tipped shape. For these parameters, the director field seeks a compromise between an undeformed, homogeneous field, and a field that is parallel to the boundary. For more extreme values of the parameter $\gamma$, we find that for $\gamma \to 0$ the director field becomes homogeneous, whereas for $\gamma \gg 1$ the director field becomes rigidly anchored to the boundary.

This behaviour is best observed by evaluating the director field at the boundary of the domain, as shown in Fig. \ref{FigBoundaryField}. Here, the value of the director field on the boundary of the domain $\Omega$ is plotted as a function of the polar angle $\varphi$ in the first quadrant, and for various values of $\gamma$ ranging between $\gamma = 0.1$ and $\gamma = 10$.
\begin{figure}[h]
\includegraphics[width=0.9\columnwidth]{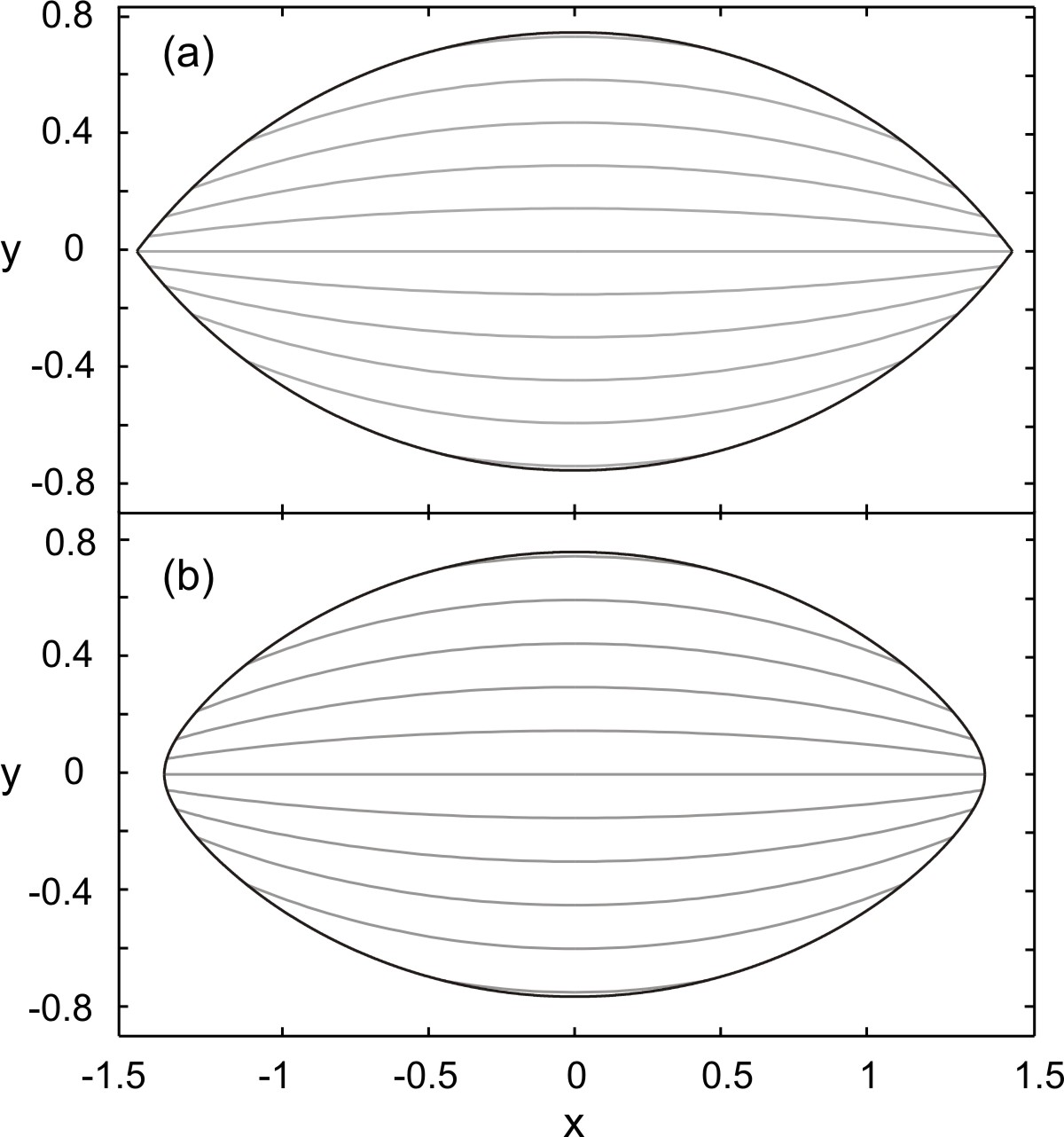}
\caption{Field lines of the optimal director field for $\gamma = 2$ and a fixed shape within the class of smooth spindle shapes: (a) sharp-tipped shape with aspect ratio $\aspectround = 2$ and $R_c = 0$, and (b) round-tipped shape with $\aspectround = 1.85$ and $R_c = 0.21$, corresponding to $\aspectsharp = 2$, and $\lambda = 0.99$.}  \label{FigFieldLines}
\end{figure}
The (red) dashed line in Fig. \ref{FigBoundaryField} indicates the limiting case of a field parallel to the boundary at all points. For sharp-tipped shapes, at $\varphi = 0$, this field goes to the value $\alpha$ as defined in Eq. (\ref{EqCuspAngle}) (see Fig. \ref{FigBoundaryField}(a)),  while for round tips the parallel field always goes to $-\pi / 2$.

\begin{figure}[h]
\includegraphics[width=0.9\columnwidth]{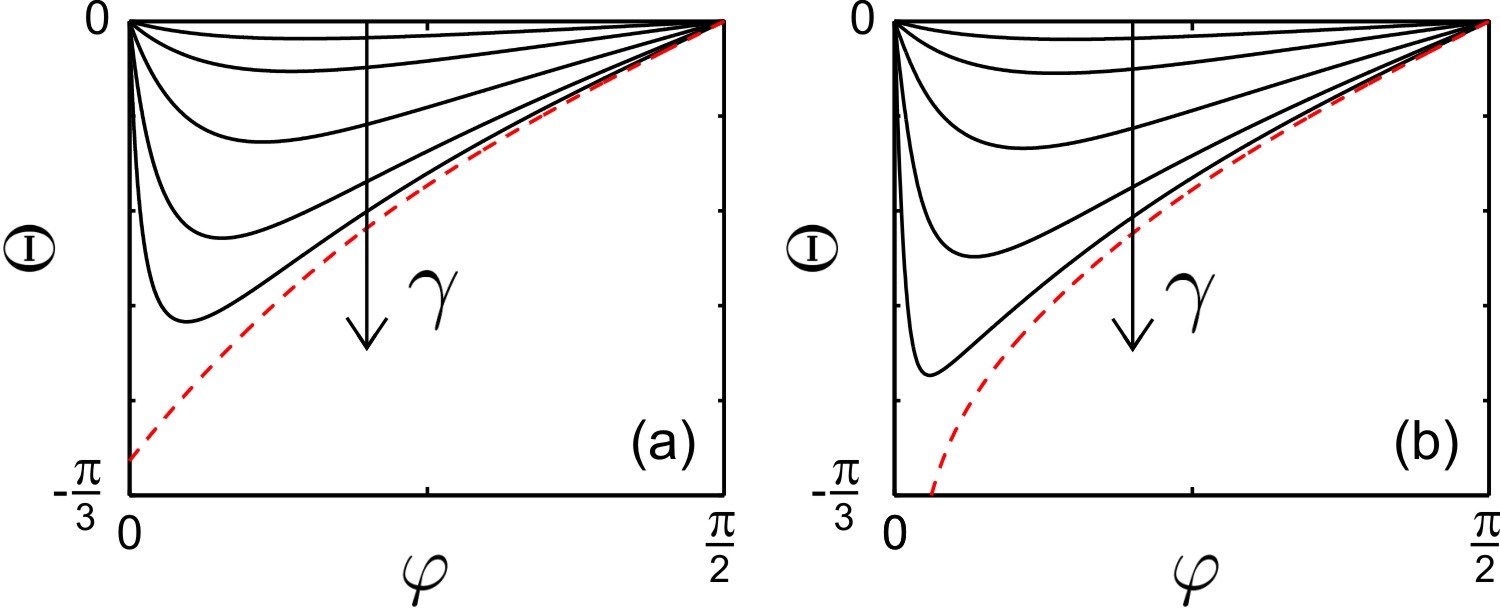}
\caption{The angle $\Theta$ of the optimal director field with the long axis of the droplet, evaluated at the boundary of the domain, and plotted as a function of the polar angle for various values of $\gamma$ ranging between $0.1$ and $10$. The arrow indicates the direction of increasing $\gamma$. Dashed (red) line indicates a director field perfectly aligned to the boundary, as would occur in the limit $\gamma\to\infty$. Shape parameters are identical to those of Fig. \ref{FigFieldLines}, with (a) a sharp-tipped shape with aspect ratio $\aspectround = 2$ and $R_c = 0$, and (b) a round-tipped shape with $\aspectround = 1.85$ and $R_c = 0.21$.}  \label{FigBoundaryField}
\end{figure}

The optimal fields for sharp-tipped shapes turn out to be very similar to the bicircular director field from Fig. \ref{FigBicircularField}. Indeed, when we restrict the director field to be bicircular and minimize the free energy with respect to position $L$ of the (virtual) boojum, we find that the optimal bicircular field is numerically indistinguishable from the calculated optimal director field. The relative differences in free energy between the optimal field and the bicircular field are shown in Figure \ref{FigDeltaF}(a), for a large range of values of $\gamma$ and aspect ratios $\aspectsharp$. There is a maximum at $\gamma \simeq 0.5$ and $\aspectsharp \simeq 1.3$, where the relative energy difference is $\sim 3.5 \cdot 10^{-4}$. For all other parameter values the bicircular field performs even better.

\begin{figure}[h]
\includegraphics[width=0.9\columnwidth]{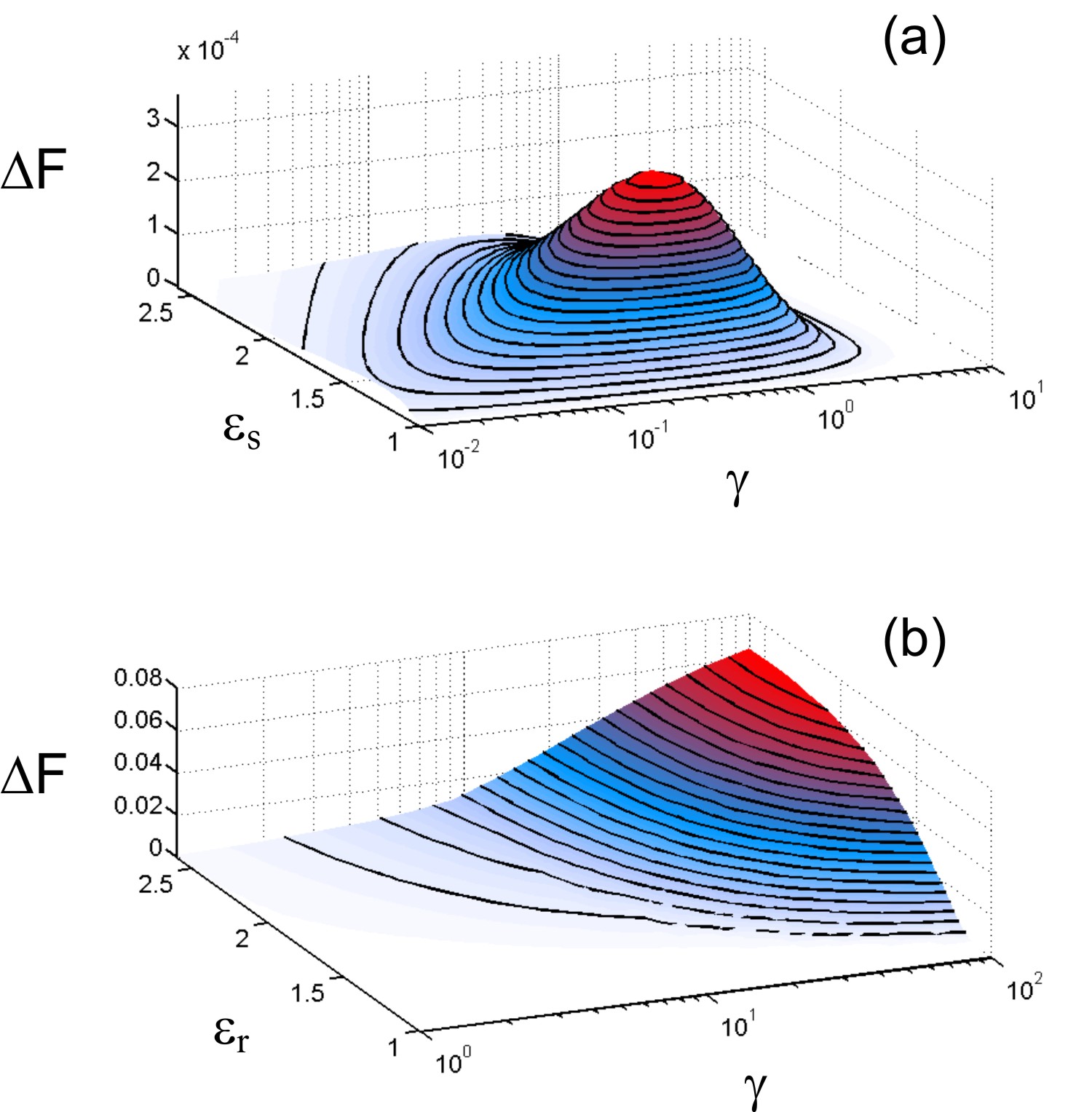}
\caption{Relative energy difference $\Delta F = (F_c - F) / F$, of the optimal director field with energy $F$ and the optimal bicircular field with energy $F_c$, shown as a function of the domain aspect ratio $\aspectsharp$, $\aspectround$, and the line tension anisotropy $\gamma$. Panel (a): sharp-tipped spindle shapes, (b): round-tipped spindle shapes with a fixed radius of curvature $R_c = 0.2$.}  \label{FigDeltaF}
\end{figure}

However, the boundary condition specified by Eq. (\ref{dTdn}) cannot be satisfied for all values of $\varphi$ for a single choice of $L$. This can be seen by solving for $L$, for particular choices of $\varphi$ for which the equations simplify, e.g., $\varphi \to 0$ and $\varphi = \pi / 4$. For these two cases, two different results for $L$ are obtained. Thus, although the likeness of the two fields is tantalizing, this proves that the bicircular field is not an exact analytical solution to the boundary value problem (\ref{Laplace}, \ref{dTdn}), on a spindle shaped domain. For all practical purposes however, the usage of the spindle field instead of the optimal solution appears justified for spindle shapes \textit{with sharp tips}. The reduction of the field optimization to a single parameter to be optimized for makes this especially convenient.

For domains with round tips the bicircular field is not nearly such a good approximation to the optimal field as it is for sharp-tipped domains. This is evidenced in Fig. \ref{FigDeltaF}(b), where the relative energy difference between the optimal field and the bicircular field is plotted against $\gamma$ and the aspect ratio $\aspectround$, while maintaining a constant radius of curvature $R_c = 0.2$. For large values of $\gamma$, the bicircular field cannot follow the domain boundary in the tip very well, leading to relatively high energy differences. But even for the rounded tips, there exist large regions in parameter space where the bicircular field approximates the optimal field very well. For smaller radii of curvature $R_c$, the bicircular field performs better, whereas for larger $R_c$ it performs worse.

Having established that the bicircular field approximates the optimal field very well for sharp-tipped shapes, we can characterize the optimal field by means of the virtual boojum locations $L$ of the bicircular field. Figure \ref{FigEpsilonL} shows a plot of the optimal value of the location $L$ of the (virtual) boojum, as a function of the shape eccentricity $\aspectsharp$ and various values of the anisotropic boundary energy strength $\gamma$. At $\aspectsharp = 1$ $L$ obeys Eq. (\ref{Ldisk}). Clearly, the higher $\gamma$, the closer the virtual boojums are located to the tip of the droplet, until in the limit $\gamma \to \infty$ the boojums are located exactly on the tips of the spindle domain.

\begin{figure}[h]
\includegraphics[width=0.85\columnwidth]{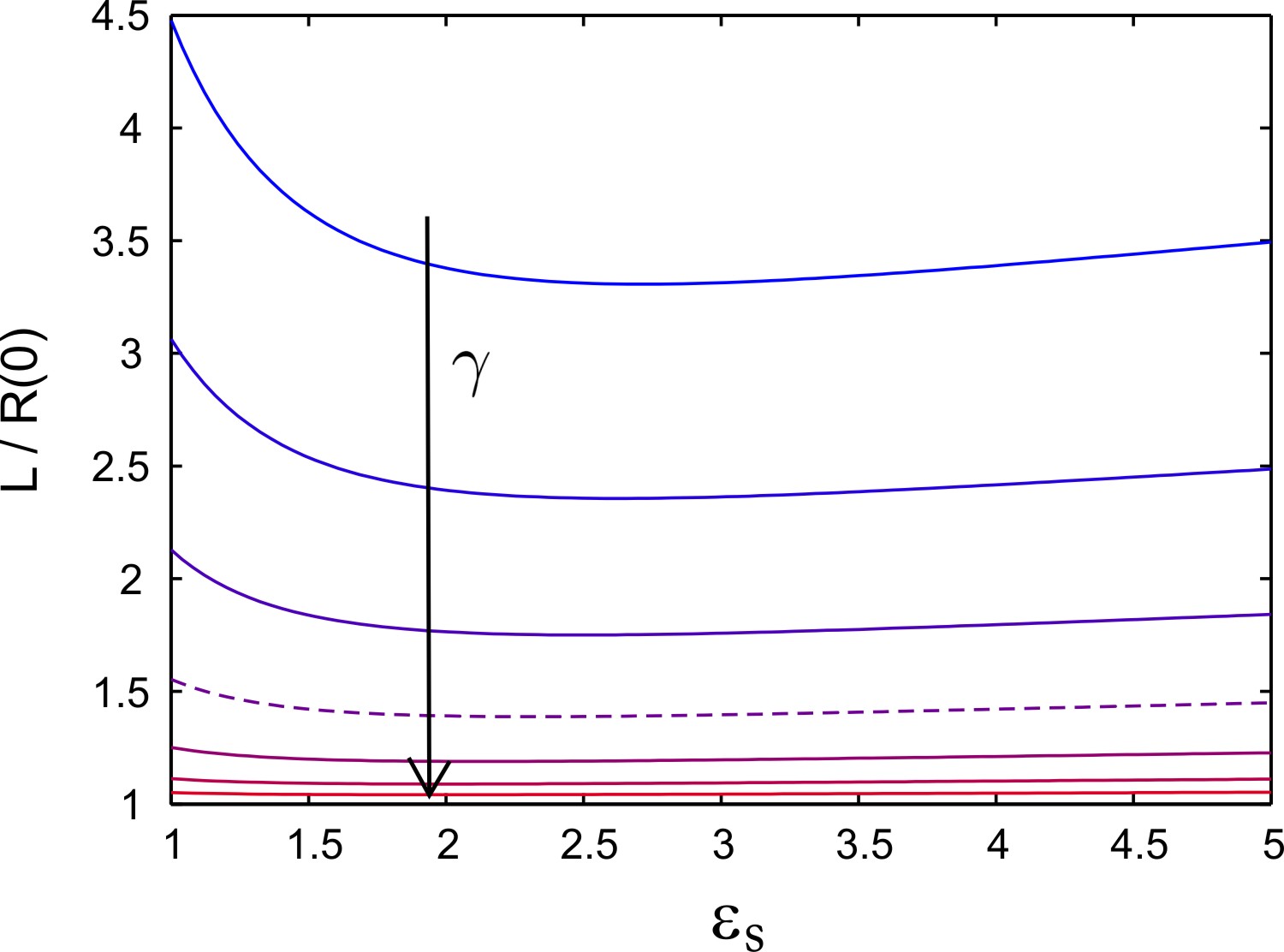}
\caption{Virtual boojum location $L$ of the optimal bicircular field, in units of the location $R(\varphi = 0)$ [see Eq. (\ref{Rphi})] of the tip of the droplet, and as a function of droplet aspect ratio $\aspectsharp$, for logarithmically spaced values of $\gamma$ between $0.1$ and $10$. The arrow indicates the direction of increasing $\gamma$, with the $\gamma = 1$ case indicated by a dashed line.}  \label{FigEpsilonL}
\end{figure}

\end{section}

\begin{section}{The optimal shape\label{SecOptimalShape}}
In this section we determine the optimal shape within the class of sharp and smooth spindle shapes by variationally minimizing the free energy, Eq. (\ref{FreeEnergy}), with respect to the shape parameters $\aspectround$ and $R_c$ given by Eqs. (\ref{epseff}) and (\ref{Rc}). At each step in the minimization procedure, we employ our solution method of Sec. \ref{SecFixedDomain} to constrain the director field to the optimal field for the current shape parameters.

Results of this optimization procedure are given in Figure \ref{FigOptShape}, showing the optimal shape for various values of $\omega$ between $0.01$ and $100$, and $0< \gamma \leq \omega$. Each line corresponds to a particular value of $\omega$, where arrows are used to indicate the direction of increasing $\omega$, and increasing $\gamma$ (for Fig. \ref{FigOptShape}(d)).
Panel (a) shows the radius of curvature of the tip of the domain, as a function of the ratio $\gamma / \omega$, which measures the anisotropy of the interfacial energy. For large enough $\gamma / \omega$, $R_c$ becomes nought, indicating a transition to a sharp tip. Panel (b) shows the tip angle of the optimal shape for the case of sharp-tipped shapes  indicated with dashed lines. The solid lines in this panel are extrapolations, and indicate tip angles if the tips were \textit{forced} to be sharp, i.e., if we ignore the possibility of the rounded tip. Panel \ref{FigOptShape}(c) shows the aspect ratio of the optimal shape, with the convention that solid lines indicate shapes with rounded tips, using the same data points as those in panel (a), whereas dashed lines indicate sharp-tipped shapes corresponding to the data points shown in panel (b).

These figures show that for $\gamma \to 0$, the boundary energy is completely determined by the isotropic line tension $\omega$ and the domain shape becomes perfectly circular in order to minimize its circumference, as evidenced by the radius of curvature $R_c \to 1$ and $\aspectround \to 1$.
As $\gamma$ is increased, the tip becomes sharper (decreasing $R_c$) while simultaneously the shape elongates (increasing $\aspectround$), until at some critical value of $\gamma < \omega$ the radius of curvature $R_c = 0$, and the shape tip becomes sharp.
It should be noted that the transition point between round and sharp tips occurs at different values of $\gamma / \omega$ when $\omega$ is varied. In the limit $\omega \to \infty$, the transition occurs immediately at $\gamma / \omega = 0$ (the optimal shape is always sharp-tipped), and the transition point moves up monotonically towards $\gamma / \omega \simeq 0.45$.
Continuing with increasing $\gamma$ beyond the transition point, the tip remains sharp and the shape will continue elongating, until in the limit $\gamma \to \omega$ the aspect ratio $\aspectround \to \infty$.

Finally, the two shape parameters for round-tipped shapes are plotted against each other in the inset Fig. \ref{FigOptShape}(d), showing clearly the tendency of increasing $\gamma$ to elongate the droplet and sharpen the tip, whereas increasing $\omega$ tends to favour more circular domain shapes. Additionally, the inset contains no data points in the upper right part of the figure. The corresponding shapes would have high aspect ratios but a radius of curvature comparable to the droplet waist ($R_c \simeq 1$), and apparently there exist no values of $\gamma$ and $\omega$ for which such shapes become energetically favorable.

\begin{figure}[t]
\includegraphics[width=0.95\columnwidth]{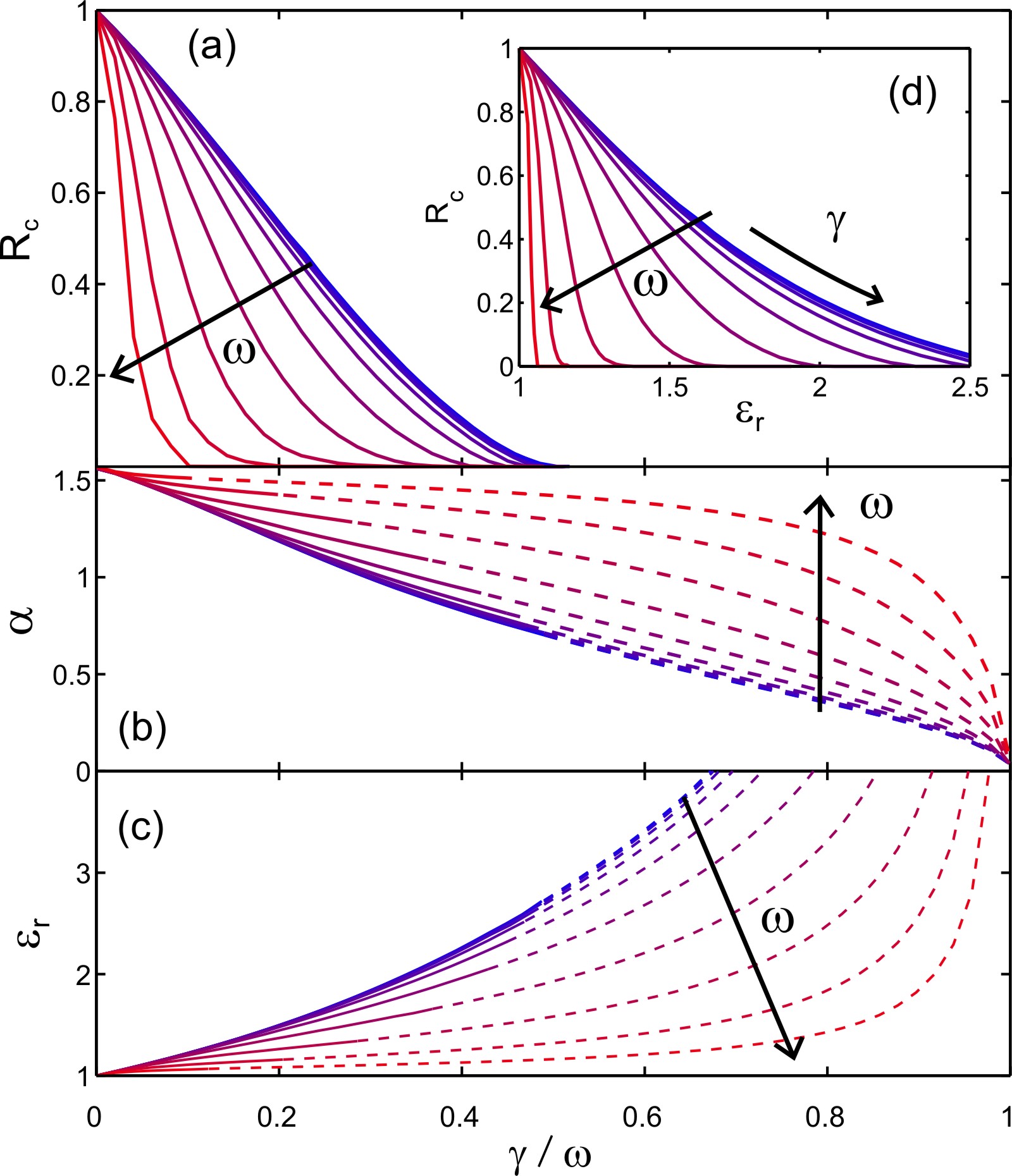}
\caption{The optimal domain shape, characterized by its aspect ratio, $\aspectround$, and radius of curvature at the tip, $R_c$, plotted for various values of the dimensionless isotropic line tension $\omega = 0.01 \ldots 100$, logarithmically spaced with the direction of increasing $\omega$ indicated with an arrow. Panel (a) shows $R_c$ as a function of the ratio of anisotropic and isotropic line tension $\gamma / \omega$, and panel (b) the tip angle of the optimal shape, as a function of $\gamma / \omega$, indicated with dashed lines. Solid lines in this panel are extrapolations that indicate the tip angle where the shape constrained to have a sharp tip. Panel (c) shows the aspect ratio $\aspectround$ of the optimal shape as a function of $\gamma / \omega$, where solid and dashed lines are used for round and sharp tips, respectively. In panel (d) $R_c$ and $\aspectround$ are plotted against each other, the arrows indicating the direction of increasing material constants $\omega$ and $\gamma$.}  \label{FigOptShape}
\end{figure}

\begin{figure}[h]
\includegraphics[width=0.95\columnwidth]{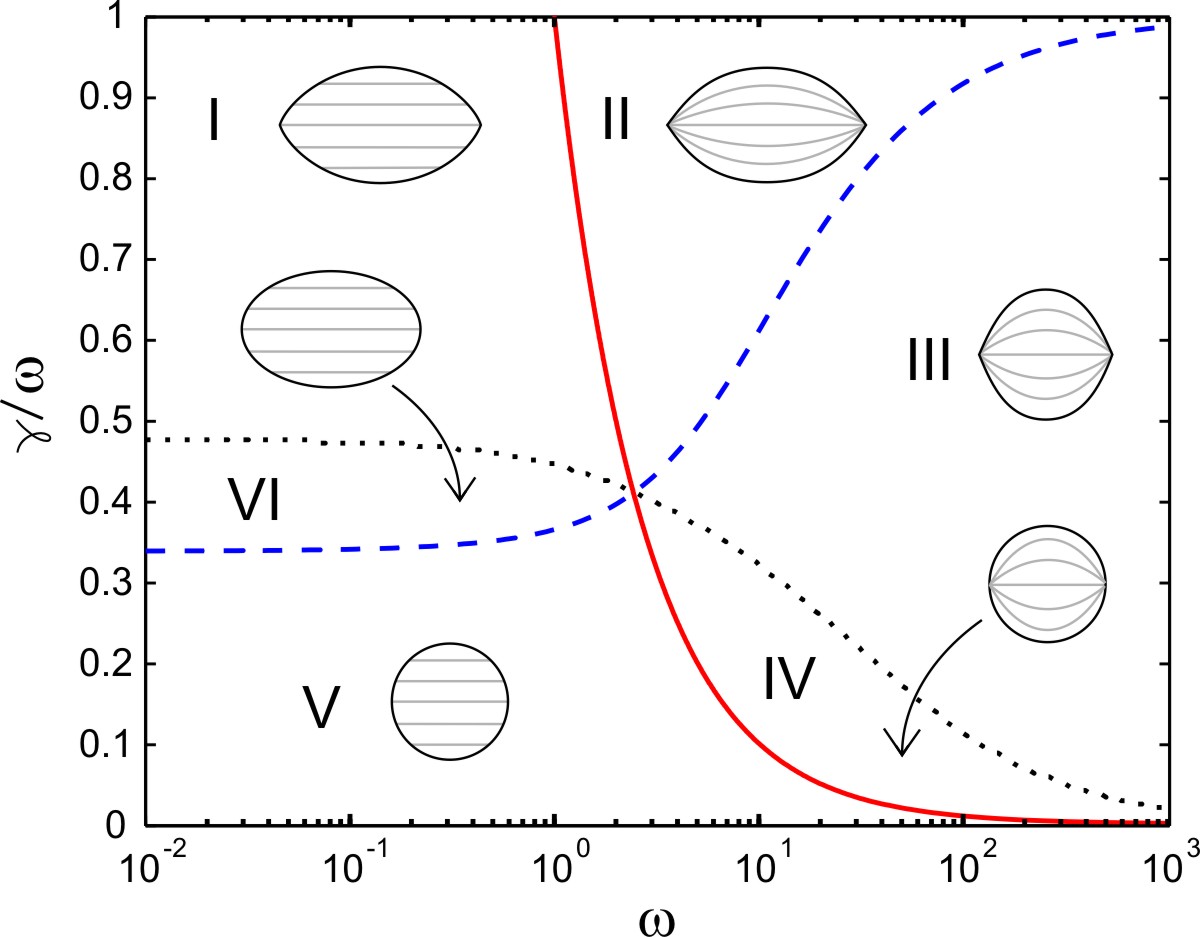}
\caption{ The shape and director field of the nematic domains are shown as a function of the dimensionless anchoring strength $\gamma/\omega$ and dimensionless surface tension $\omega$. Six regions, indicated by the Roman numerals, are found that follow from three characteristics we distinguish. The dotted line gives the boundary between a smooth and a sharp tip of the domain; the dashed line shows the transition from an elongated to a round domain; the solid line is the boundary between a uniform and a bipolar director field. \label{FigPhaseDiag}}
\end{figure}

The results of our numerical calculations of the optimal director-field structure and  shape (within our class of smooth spindle shapes), can be gathered in a single phase diagram as shown in Figure \ref{FigPhaseDiag}. We distinguish six regimes, labeled I through VI, based on three defining properties:

\textit{i) The director field being homogeneous or curved.} For small values of $\gamma$, the bulk elasticity dominates over the interfacial anchoring, so the director field is rigid and homogeneous. For large values of $\gamma$, the anisotropic boundary energy dominates and the director field will accommodate by aligning itself parallel to the shape boundary. The (smooth) transition between the two types of field is chosen to occur at $\gamma = 1$, and is indicated by the solid (red) line in Fig. \ref{FigPhaseDiag}. Points to the left of this line correspond to homogeneous director fields, whereas points to the right have a bipolar (curved) director field. The crossover value $\gamma=1$ is sensible because the anchoring energy and the elastic deformation energy are then approximately equal.

\textit{ii) The droplet tip being rounded or sharp.} For small values of the dimensionless anisotropic line tension $\gamma$, the dimensionless isotropic line tension $\omega$ dominates the total interfacial (line) free energy. The shorter total boundary length associated with rounded tips outweighs the rise in anisotropic surface energy due to the larger attack angles at such a round tip. Hence, for small values of $\gamma$ the droplet tip becomes rounded, whereas for higher values of $\gamma$ the opposite situation occurs and the tip becomes sharp. In Fig. \ref{FigPhaseDiag} the transition from round to sharp is indicated by a black, dotted line. Points above this line have sharp tips, points below have rounded tips.

\textit{iii) The elongation of the droplet.} The elongation of a droplet is defined by its aspect ratio $\aspectround$. [Note that $\aspectround = \aspectsharp$ for sharp-tipped shapes.] For high isotropic line tensions $\omega$, the droplets will tend to be close to circular to minimize their interfacial energy. If the anisotropic line tension $\gamma$ increases and approaches $\omega$, the director field aligns along the boundary. If the isotropic line tension is small, the director field is rigid. Aligning the field by elongating the droplet becomes energetically favourable over deforming the director field, at the expense of increasing the boundary length and associated isotropic interfacial energy. In Fig. \ref{FigPhaseDiag}, the dashed (blue) line marks points with constant aspect ratio, in this case equal to $\aspectround = 2$. All points located above this line have larger aspect ratios and are more elongated; all points below it have lower aspect ratios and are more round.

In principle, these three properties allow for eight distinct topological combinations in the phase diagram. However, we only distinguish six possible regimes in the phase diagram, defined by the limiting behavior of the dividing lines. These six regimes extend all the way to the extreme values of $\gamma / \omega$, and $\log\omega \to \pm \infty$. These six regimes are labeled I through VI in fig. \ref{FigPhaseDiag}, and are defined by the following characteristics:
I) elongated domain, with sharp tip and homogeneous director field, II) elongated domain, with sharp tip and curved director field, III) round domain, with sharp tip, and curved director field, IV) round domain, with round tip, and curved director field, V) round domain, with round tip, and homogeneous director field, and finally, VI) elongated domain, with round tip, and homogeneous director field.

The remaining two topological combinations are elongated round-tipped domains with a curved director field, and round droplets with a sharp tip and homogeneous director field. These regimes would constitute a finite domain in the center of the phase diagram. However, the lines defining transitions between homogeneous or curved director fields (i), and round and elongated shapes (iii), are somewhat arbitrary as these transitions are smooth and the location of the transition is a matter of definition only. By choosing different values for the critical parameters, these lines can be moved and the phase diagram will look differently in the center region. Note that the transition between round and sharp tipped shapes is sharp and resembles a second order phase transition.

The current critical values are chosen such that the dividing lines all cross in a single point. Moving one of the lines will immediately introduce one of the remaining two topological combinations. But these regions typically would only occupy a small area of the phase diagram and are subject to arbitrary criteria. By picking the current values, the phase diagram only shows the essential topological combinations that would be always present as a limiting case for any reasonable choice of critical values.

If we compare the state diagram Fig. \ref{FigPhaseDiag} for two dimensional nematic domains with that calculated by Prinsen et al. \cite{prinsen04} for three dimensional nematic drops, we find that they are remarkably similar. Our phase diagram has one extra attribute not present in that of Prinsen et al. \cite{prinsen04}, being the transition from round to sharp tipped shapes. This generates two more regimes not present in the earlier work: regimes IV and VI, i.e., rounded shapes with curved director fields, and elongated domains with round tips and uniform director field. Interestingly, the crossover line for the transition between elongated and round shapes coincides to within numerical precision, even within regime VI. 

An important point to note is that for a homogeneous director field, i.e., in the limit $\omega \to 0$, the optimal shape can be obtained from the so-called Wulff construction \cite{wulff01}. This shape is then completely determined by the anisotropy $\gamma/\omega$ of the interfacial tension.
However, the Wulff shape is not part of our class of spindle shapes and, in particular, the limiting value of the sharp-round transition in Fig. \ref{FigPhaseDiag} of $\gamma / \omega \simeq 0.45$ does not occur at the value of $\gamma / \omega = 1/3$ as predicted by the Wulff construction.

The fact that the Wulff shape is not included in the spindle shape class can be seen as follows: the Wulff shape found just above the transition point has an aspect ratio of $2$, while the tip angle $\alpha_{\rm wulff} \to \pi / 2$. A sharp-tipped spindle shape of aspect ratio $2$ would necessarily have a tip angle $\alpha \simeq 0.3 \pi $, as dictated by Eq. (\ref{EqCuspAngle}).
However, the spindle shapes we find for a uniform director field have a free energy that is at most $0.1 \%$ larger than that of the Wulff shape (at least for the equivalent shapes in three dimensions \cite{prinsen03}).
In the discussion in the next section we will elaborate on how the smooth spindle class can be expanded to include shapes with arbitrary aspect ratios and tip angles, and approximate the Wulff shape arbitrarily closely.

\end{section}

\begin{section}{Conclusions and Discussion\label{SecConclusions}}
In this paper we have presented a general approach to find the optimal shape and director field of two-dimensional nematic domains. Using conformal mappings, we can transform the original boundary value problem with nonlinear boundary conditions on the general domain $\Omega$, to a boundary value problem on the unit disk. Next, by noting that the value of the director field on the boundary uniquely determines the solution, we can rewrite the problem in terms of the Fourier coefficients of the value of the director field on the boundary of the unit disk. Finally, we numerically solve for these coefficients with an iterative procedure, which is easily implemented and can be done at relatively low computational cost due to the usage of Fast Fourier Transforms.

Using this methodology, we investigate the optimal field within a realistic class of spindle shapes, which can have sharp as well as round tips. We find that the optimal director field for spindle-shaped domains with a sharp tip is numerically equivalent to a field that consists of circle sections. This means that for practical purposes such a bicircular field can be used to model the director field of spindle shaped domains of particles that prefer planar alignment of the director field to the boundary.

By establishing the connection of the optimal director field to a given shape, we can search for the optimal shape within our class of smooth spindle shapes for given material constants. The results for the optimal shape are compiled into a single phase diagram (Fig. \ref{FigPhaseDiag}), in which we distinguish six characteristic regimes of domain shape and director field. The optimal shape and fields are classified based on whether the domains are elongated or round, have a sharp or a smooth tip, and have a bipolar or a homogeneous director field. We find that if the anisotropy of the interfacial energy is small, the domain has a round tip. 

Although we have limited ourselves in this work to realistic shapes based on domains freely floating in an isotropic background, our approach is straightforwardly extended to cover several other interesting related systems. For instance, our methodology is applicable to any shape for which the conformal mapping to a circle is known. A possible application of this is, e.g., the study of nematics in confined geometries, or the coalescence of two drops by taking a peanut-shaped domain. This latter domain can for instance easily be constructed by a conformal map of the form $f(z) \propto \tan(\lambda z)$, with $\lambda$ a parameter between $0$ and $1$. Alternatively, the class of smooth spindle shapes can be easily extended to a more general class.

We have only considered a preferred planar alignment of the director field to the domain surface. However, the preferred type of anchoring could be perpendicular to the boundary (i.e., homeotropic). In that case $a_2 < 0$ in our boundary energy density Eq. (\ref{BoundaryEnergya0a2}), and typically we expect a radial director field with a hedgehog defect at the center of the domain, at least for large enough domains \cite{verhoeff11lm}.

The defect free case with homeotropic anchoring, i.e., for small enough domains, is readily covered by our method as briefly explained at the end of section \ref{SecFixedDomain}. More generally speaking, our calculations make no explicit use of the functional form of the boundary energy. They could therefore identically be applied to any other type of boundary-energy prescription.

To account for defects in the interior of the domain, our method can be extended because there exist conformal maps that map between the unit disk and a disk with a small hole cut out at an arbitrary location \cite{Kober}. These could then be used to generate domains with vanishingly small holes, and by setting Dirichlet boundary conditions at the boundary of such a hole it would be possible to emulate the presence of a defect in the field.

Unfortunately, there are a number of caveats associated with our methodology.
Firstly, the crowding phenomenon of the conformal mapping, as discussed at the end of Sec. \ref{SecSpindle}, makes it computationally difficult to accurately calculate the director field for domains with high aspect ratios larger than approximately 3, and with very sharp tips.

Secondly, we found that the optimal shape for homogeneous director fields, as given by the Wulff construction, possesses particular combinations of aspect ratio and tip angle that are not supported by our smooth spindle shape class. However, this caveat can be circumvented by sequentially applying more than one conformal map of a unit disk to a spindle domain. This way, it is possible to generate spindle-like domains, but with arbitrary aspect ratios and arbitrary tip angles, including smooth tips.

Thirdly and finally, the method as presented in this paper works only in the equal-constant approximation, because only then the boundary value problem defining the director field simplifies to a Neumann problem. Likewise, similar issues arise in generalisations to three dimensions, even if we invoke cylindrical symmetry. This means that for these cases the method cannot be \textit{directly} applied, but needs to be modified. We leave this for future work.

\end{section}

\begin{acknowledgments}
The work of R.O. forms part of the research programme of the Dutch Polymer Institute (DPI, project 648).
\end{acknowledgments}


\begin{thebibliography}{99}
\bibitem{degennes93} de~Gennes,~P.~G.; Prost,~J. \emph{The Physics of Liquid Crystals}; \newblock Oxford University Press, 1993.
\bibitem{prinsen03} Prinsen,~P.; van~der Schoot,~P. \emph{Physical Review E} \textbf{2003}, \emph{68}, 021701.
\bibitem{prinsen04} Prinsen,~P.; van~der Schoot,~P. \emph{European Physical Journal E}  \textbf{2004}, \emph{13}, 35--41.
\bibitem{fraden85} Fraden,~S.; Hurd,~A.~J.; Meyer,~R.~B.; Cahoon,~M.; Casper,~D.~L. \emph{J. Phys. (Paris), Colloq.} \textbf{1985}, \emph{46}, C3--85.
\bibitem{drzaic95} Drzaic,~P. \emph{Liquid Crystal Dispersions}; \newblock World Scientific, Singapore, 1995.
\bibitem{tang95} Tang,~J.~X.; Fraden,~S. \emph{Liquid Crystals} \textbf{1995}, \emph{19},   459--467.
\bibitem{riviere95} Rivi\`{e}re,~S.; Meunier,~J. \emph{Physical Review Letters} \textbf{1995}, \emph{74}, 2495--2498.
\bibitem{rapini69} Rapini,~A.; Papoular,~M. \emph{J. Phys. (Paris), Colloq.} \textbf{1969},   \emph{30}, C4--54.
\bibitem{burylov97} Burylov,~S.~V. \emph{Journal of Experimental and Theoretical Physics} \textbf{1997}, 85, 873--886.
\bibitem{heras09} de las Heras,~D.; Velasco,~V.; Mederos,~L.; \emph{Physical Review  E} \textbf{2009}, \emph{79}, 061703.
\bibitem{ondris93} Ondris-Crawford,~R.~J.; Crawford,~G.~P.; \v{Z}umer,~S.; Doane,~J.~W. \emph{Physical Review Letters} \textbf{1993}, \emph{70}, 194--197.
\bibitem{verhoeff09} Verhoeff,~A.~A.; Otten,~R.~H.~J.; van der Schoot,~P.; Lekkerkerker,~H.~N.~W.;
\emph{Journal of Physical Chemistry B} \textbf{2009}, \emph{113}, 3704-3708.
\bibitem{verhoeff11jcp}  A. A. Verhoeff, R. H. J. Otten, P. van der Schoot, H. N. W. Lekkerkerker, Journal of Chemical Physics \textbf{2011}, \emph{134}, 044904.
\bibitem{verhoeff11lm}  A. A. Verhoeff, R. H. J. Otten, I. A. Bakelaar, P. van der Schoot, H. N. W. Lekkerkerker, Langmuir \textbf{2011}, \emph{27}, 116.
\bibitem{rudnick95} Rudnick,~J.; Bruinsma,~R. \emph{Physical Review Letters} \textbf{1995}, \emph{74}, 2491--2494.
\bibitem{fang97} 7.J. Fang, E. Teer, C. M. Knobler, K.-K. Loh, and J. Rudnick, Phys. Rev. E 56, 1859 (1997) 
\bibitem{rudnick99} Rudnick,~J.; Loh,~K.-K.; \emph{Physical Review E} \textbf{1999}, \emph{60}, 3045--3062.
\bibitem{zocher29} Zocher,~H.; Jacobsohn,~K. \emph{Kolloidchemische Beihefte} \textbf{1929}, \emph{28}, 167--206.
\bibitem{kaznacheev02} Kaznacheev,~A.~V.; Bogdanov,~M.~M.; Taraskin,~S.~A. \emph{Journal of Experimental and Theoretical Physics} \textbf{2002}, \emph{95}, 57--63.
\bibitem{kaznacheev03} A. V. Kaznacheev, M. M. Bogdanov and A. S. Sonin,  J. Exp. Theor. Phys. 97, 1159 (2003).


\bibitem{frank58} F. C. Frank, \emph{Discussions of the Faraday Society} \textbf{25}, 19--28 (1958).


\bibitem{kooij98} van~der Kooij,~F.~M.; Lekkerkerker,~H. N.~W. \emph{Journal of Physical Chemistry B} \textbf{1998}, \emph{102}, 7829--7832.
\bibitem{fischer94} Fischer,~T.~M.; Rudnick,~J.; Bruinsma,~R. \emph{Physical Review  E} \textbf{1994}, \emph{50}, 413--428.
\bibitem{loh98} Loh,~K.-K.; Rudnick,~J.; \emph{Physical Review Letters} \textbf{1998}, \emph{81}, 4935--4938.
\bibitem{loh00} Loh,~K.-K.; Rudnick,~J.; \emph{Physical Review E} \textbf{2000}, \emph{62}, 2416--2427.

\bibitem{pettey99} Pettey,~D.; Lubensky,~T.~C. \emph{Physical Review  E} \textbf{1999}, \emph{59}, 1834--1845.

\bibitem{schwartz94} Schwartz,~D.~K.; Tsao,~M.-W.; Knobler,~C.~M. \emph{Journal of Chemical Physics} \textbf{1994}, \emph{101}, 8258--8261.
\bibitem{dolganov07} Dolganov,~P.~V.; Nguyen,~H.~T.; Joly,~G.; Dolganov,~V.~K.; Cluzeau,~P. \emph{European Physics Letters} \textbf{2007}, \emph{78}, 66001.
\bibitem{Burylov97} S.~V.~Burylov, Zh. Eksp. Teor. Fiz. \textbf{112}, 1603 (1997)
\bibitem{davidson11} A. J. Davidson and N. J. Mottram, Euro. Jnl of Applied Mathematics (2012), vol. 23, pp 99-119.   



\bibitem{williams86} Williams,~R.~D. \emph{J. Phys. A: Math. Gen.} \textbf{1986}, \textbf{19}, 3211--3222.

\bibitem{bernal41} Bernal,~J.~D.; Fankuchen,~I.; \emph{Journal of General Physiology},   \textbf{1941}, \emph{25}, 111.
\bibitem{wulff01} Wulff,~G.; \emph{Z. Kristallogr.} \textbf{1901},   \emph{34}, 449.


\bibitem{dogic01} Dogic,~Z.; Fraden,~S. 
\emph{Philosophical Transactions of the Royal Society of   London Series a-Mathematical Physical and Engineering Sciences}
    \textbf{2001}, \emph{359},   997--1014.
\bibitem{chen96} Chen,~W.~L.; Sato,~T.; Teramoto,~A. \emph{Macromolecules} \textbf{1996},   \emph{29}, 4283--4286.

\bibitem{scharkowski93} Scharkowski,~A.; Crawford,~G.~P.; $\breve{{\rm Z}}$umer,~S.; Doane,~J.~W. \emph{Journal of Applied Physics} \textbf{1993}, \emph{73}, 7280--7287.
\bibitem{vertogen88} Vertogen,~G; de Jeu,~W.~H. \emph{Thermotropic Liquid Crystals, Fundamentals}, Springer-Verlag, Berlin, 1988.
\bibitem{vdschoot99} van der Schoot,~P.; \emph{J. Phys. Chem. B} \textbf{1999}, 103, 8804-8808
\bibitem{kalugin98} Kalugin, ~A.~G.; Golubyatnikov,~A.~N. \emph{Proc. Steklov Inst. Math.} \textbf{1998}, \emph{223}, 168--174.
\bibitem{Gradshteyn}  Gradshteyn,~L.~S.; Ryzhik,~I.~M.; eds., \emph{Table of Integrals, Series, and Products} (Academic Press, San Diego, 2000), sixth  ed.
\bibitem{Brigham}  Brigham,~E.~O.; The fast Fourier transform and its applications, Prentice Hall International, London, 1988.
\bibitem{BenderOrszag} Bender,~C. M.; Orszag,~S.~A.; Advanced Mathematical Methods for Scientists and Engineers: Asymptotic Methods and Perturbation Theory (Springer Verlag, New York, 1999).
\bibitem{Nehari}  Nehari,~Z.; Conformal mapping, McGraw-Hill, London, \textbf{1952}.
\bibitem{Roach} G. F. Roach, Green's functions, Cambridge University Press, Second edition \textbf{1982}.
\bibitem{puech10} Puech,~N.; Grelet,~E.; Poulin,~P.; Blanc,~C.; van der Schoot,~P.; Phys. Rev. E \textbf{2010}, \emph{82}, 020702.
\bibitem{zhang06} Zhang,~Z. X.; van Duijneveldt,~J.~S.; J. Chem. Phys. \textbf{2006}, \emph{124}, 154910.
\bibitem{Henrici} Henrici,~P.; Applied and Computational Complex Analysis, Vol. III, John Wiley \& Sons \textbf{1986}.
\bibitem{Churchill} Churchill,~R.~V.; Complex Variables and Applications, Second Edition, McGraw-Hill Book Company, Inc. \textbf{1960}.
\bibitem{Dettman}  Dettman,~J.~W.; Applied Complex Variables, The MacMillan Company \textbf{1965}.
\bibitem{Wegmann} Wegmann,~R.; Constr. Approx. 10, pp. 179-186 \textbf{1994}.
\bibitem{DeLillo} DeLillo,~T.~K.; SIAM J. Numer. Anal. 31(3), pp. 788-812 \textbf{1994}.
\bibitem{Zemach} Zemach,~C.; Journal of Computational and Applied Mathematics 14 (1-2), pp. 207-215.
\bibitem{Kober} Kober,~H.; Dictionary of Conformal Representations, New York: Dover \textbf{1957}.
\end{thebibliography}
\end{document}